\documentclass[aps,prb,twocolumn,superscriptaddress,nofootinbib,longbibliography,floatfix]{revtex4-2}
\usepackage{amsmath,amssymb,amsfonts,mathtools}
\usepackage{amsthm}
\usepackage{braket}
\usepackage{graphicx}
\usepackage{microtype}
\usepackage[colorlinks=true,linkcolor=blue!60!black,citecolor=blue!60!black,urlcolor=blue!60!black]{hyperref}
\usepackage{xcolor}
\usepackage{tikz}
\usetikzlibrary{positioning,arrows.meta,calc}

\newtheorem{proposition}{Proposition}
\newtheorem{lemma}{Lemma}
\newtheorem{corollary}{Corollary}

\newcommand{\XN}{\mathcal{X}_N}
\newcommand{\KL}[2]{D_{\mathrm{KL}}\!\left(#1\,\middle\|\,#2\right)}
\newcommand{\Bern}{\mathrm{Bern}}
\newcommand{\E}{\mathbb{E}}

\newcommand{\ctx}{C_{i,R}}
\newcommand{\dpast}{D_{i,R}}

\begin{document}

%\title{Finite-Sample Model Resolution for Quantum Born Distributions}
\title{Sample Complexity for Generative Learning of Quantum Measurement Data}
\author{Shuhan Zhang}
\affiliation{Department of Physics, University of California, San Diego, La Jolla, CA 92093, USA}
\author{Yi-Zhuang You}
\affiliation{Department of Physics, University of California, San Diego, La Jolla, CA 92093, USA}
\affiliation{EdenCode Inc., San Jose, CA 95134, USA}
\date{\today}

\begin{abstract}
Generative learning of a quantum measurement distribution from finite data requires choosing a model capacity that balances representation accuracy against estimation cost. We develop a bias–variance framework that decomposes the expected population Kullback–Leibler error at model capacity $s$ and sample size $N$ as $\mathcal E(s,N)=B(s)+V(s,N)$, with representation bias $B$ and a variance term $V$ capturing the mean estimation error. In the well-sampled regime, $V(s,N)\simeq[NC(s)]^{-1}$, and exponential or algebraic laws for $B$ and $C$ yield four scaling templates for the optimal error, from inverse powers of $N$ to inverse powers of $\ln N$. For local autoregressive models, the context radius sets the model capacity, and $B$ is the summed conditional mutual information lost by truncating the context. For a smoothed histogram learner, we derive an exact finite-sample estimation cost that accounts for rarely observed contexts and recovers the parameter-count law in the well-sampled regime. For transverse-field Ising Born distributions, the independently evaluated bias and estimation cost predict both the optimal context radius and the optimal error. In the well-sampled regime, the estimation cost grows exponentially with context radius. Its competition with approximately exponential bias decay in the gapped regimes predicts power-law error decay in $N$, while algebraic bias decay at criticality predicts asymptotic inverse-power decay in $\ln N$. Bulk-site results are consistent with the corresponding finite-sample predictions. These results connect the structure of quantum measurement distributions to the model capacity justified by finite data.
\end{abstract}

\maketitle

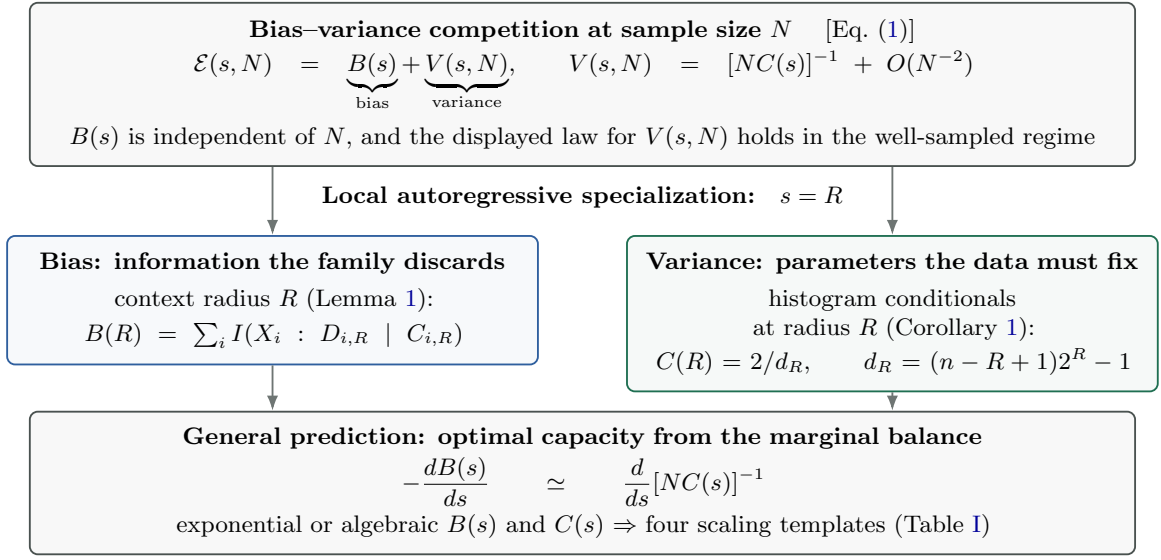
\begin{figure*}[t]
\centering
\definecolor{rmblue}{HTML}{2F5F9E}
\definecolor{rmgreen}{HTML}{1F6F52}
\definecolor{rmgray}{HTML}{4A5250}

\begin{tikzpicture}[
 every node/.style={
   align=center,
   inner sep=6pt,
   font=\fontsize{9}{11}\selectfont
 },
 box/.style={
   draw=#1,
   fill=#1!4,
   rounded corners=3pt,
   line width=0.7pt
 },
 arr/.style={
   -{Latex[length=2mm]},
   line width=0.7pt,
   rmgray!80
 }]

% General framework
\node[
 box=rmgray,
 text width=14.2cm,
 minimum height=1.0cm
] (law) at (0,0)
 {\textbf{Bias--variance competition at sample size $N$}
  \quad[Eq.~\eqref{eq:law}]\\[2pt]
  $\mathcal E(s,N)
   =\underbrace{B(s)}_{\text{bias}}
   +\underbrace{V(s,N)}_{\text{variance}},
   \qquad
   V(s,N)=[NC(s)]^{-1}+O(N^{-2})$\\[5pt]
  $B(s)$ is independent of $N$, and the displayed law for
  $V(s,N)$ holds in the well-sampled regime};

% Explicit specialization from generic capacity s to context radius R
\node[
 font=\fontsize{9}{11}\selectfont
] (specialization) at ($(law.south)+(0,-0.40)$)
 {\textbf{Local autoregressive specialization:}\quad $s=R$};

% Bias and variance for the local autoregressive model
\node[
 box=rmblue,
 text width=6.6cm,
 minimum height=1.7cm,
 anchor=north
] (bias) at ($(law.south)+(-4.1,-0.90)$)
 {\textbf{Bias: information the family discards}\\[3pt]
  context radius $R$ (Lemma~\ref{lem:cmi}):\\[2pt]
  $B(R)=\sum_i I(X_i:\dpast\mid\ctx)$};

\node[
 box=rmgreen,
 text width=6.6cm,
 minimum height=1.7cm,
 anchor=north
] (var) at ($(law.south)+(4.1,-0.90)$)
 {\textbf{Variance: parameters the data must fix}\\[3pt]
  histogram conditionals at radius $R$
  (Corollary~\ref{cor:dR}):\\[2pt]
  $C(R)=2/d_R,\qquad d_R=(n-R+1)2^R-1$};

% Optimization and scaling
\node[
 box=rmgray,
 text width=14.2cm,
 minimum height=1.0cm,
 anchor=north
] (opt) at ($(bias.south)+(4.1,-0.60)$)
 {\textbf{General prediction: optimal capacity from the marginal balance}\\[2pt]
  $\displaystyle
  -\frac{dB(s)}{ds}
  \simeq
  \frac{d}{ds}[NC(s)]^{-1}$\\[2pt]
  exponential or algebraic $B(s)$ and $C(s)$
  $\Rightarrow$ four scaling templates
  (Table~\ref{tab:templates})};

% Arrows
\draw[arr]
  (law.south -| bias.north) -- (bias.north);
\draw[arr]
  (law.south -| var.north) -- (var.north);
\draw[arr]
  (bias.south) -- (bias.south |- opt.north);
\draw[arr]
  (var.south) -- (var.south |- opt.north);

\end{tikzpicture}

\caption{The organizing principle of the paper. The target
distribution, model family, and learner determine $B(s)$ and $C(s)$, whose marginal balance selects the optimal capacity and yields the four scaling templates. For local autoregression, the generic capacity $s$ is the context radius $R$. The bias is determined by discarded conditional information, while the leading asymptotic estimation cost is determined by the parameter count. }
\label{fig:roadmap}
\end{figure*}
\section{Introduction}
\label{sec:introduction}

Repeated measurements of a quantum many-body state produce a classical distribution whose structure reflects both the state and the measurement basis. Generative models can learn these distributions and provide compact representations of quantum measurement data~\cite{Torlai2018,Carrasquilla2019,Rocchetto2018,Quek2021,Sharir2020,HibatAllah2020}. The error of such a model decreases with the number of measurement records, but the form of the decrease varies in different settings. Neural scaling studies report power-law decreases in test loss with training sample size~\cite{Hestness2017,Kaplan2020,Bahri2024}. By contrast, Hou et al.~\cite{Hou2025_MLManyMeasurements} trained attention-based networks on cluster-state measurement outcomes from Google’s Sycamore and Willow processors and reported only logarithmic decay of the relative entropy between the measured and learned distributions. Studies of quantum
measurement data also report sharp changes in learnability across physical
regimes~\cite{Hou2025_MLManyMeasurements,Kumar2026_UnlearnablePhases} and information-theoretic
scaling laws for neural quantum states~\cite{Lu2026_InfoScalingNQS}. The central question is therefore how much model capacity the available measurement records justify, and why the answer leads to such different scalings in different settings.

\begin{table}[t]
\caption{Scaling templates from the marginal balance in Eq.~\eqref{eq:balance}, with positive
exponents and $N$-independent amplitudes. Rows I--III give $s^\star$ to additive $O(1)$, and
$\asymp$ denotes equality up to bounded positive factors, so the capacity and error columns
have different stated precisions (Appendix~\ref{app:templates}). Local autoregression with
exponential bias gives row I, and algebraic critical bias gives row III.}
\label{tab:templates}
\setlength{\tabcolsep}{3pt}
\begin{ruledtabular}
\begin{tabular}{clccc}
 & $B(s)$ & $C(s)$ & $s^\star(N)$ & $\mathcal E^\star(N)$\\
\colrule
I\rule{0pt}{4.2ex} & $e^{-\alpha s}$ & $e^{-\beta s}$ & $\dfrac{\ln N}{\alpha+\beta}$ & $\asymp N^{-\alpha/(\alpha+\beta)}$\\[7pt]
II & $e^{-\alpha s}$ & $s^{-\nu}$ & $\dfrac{\ln N-(\nu-1)\ln\ln N}{\alpha}$ & $\asymp(\ln N)^\nu/N$\\[7pt]
III & $s^{-\mu}$ & $e^{-\beta s}$ & $\dfrac{\ln N-(\mu+1)\ln\ln N}{\beta}$ & $\asymp(\ln N)^{-\mu}$\\[7pt]
IV & $s^{-\mu}$ & $s^{-\nu}$ & $\asymp N^{1/(\mu+\nu)}$ & $\asymp N^{-\mu/(\mu+\nu)}$\\
\end{tabular}
\end{ruledtabular}
\end{table}

This paper argues that the competition between bias and variance~\cite{Geman1992} in a model
family of limited capacity organizes these distinct behaviors. Every generative model is
restricted by a capacity parameter $s$, be it a context radius, a bond dimension, or a
receptive field. Restricting capacity introduces a representation bias $B(s)$ but reduces the estimation cost associated with learning from finite data. We denote this cost by $V(s,N)$, the mean excess error above the best distribution in the family, which plays the role of variance. For the
model--learner pairs considered here, at fixed capacity the expected population error has the
large-$N$ asymptotic expansion
\begin{equation}
\begin{aligned}
\mathcal E(s,N)&=B(s)+V(s,N),\\
V(s,N)&=\frac{1}{N\,C(s)}+O(N^{-2}),
\end{aligned}
\label{eq:law}
\end{equation}
where the learning coefficient $C(s)$ depends on the target distribution, model family, and learner. The leading term is accurate in the \emph{well-sampled regime}, where every retained parameter has been observed many times (Sec.~\ref{sec:finiteN}). When increasing $s$ lowers the bias but raises the estimation cost, their competition determines the optimal capacity $s^\star(N)$ and error $\mathcal E^\star(N)$.  Exponential or algebraic decay of $B$ and $C$ then produces optimal errors that decay as powers of $N$, up to logarithmic factors, or as inverse powers of $\ln N$ (Fig.~\ref{fig:roadmap} and Table~\ref{tab:templates}). 

Applying the framework requires the bias and estimation cost of a given model family, learner, and target distribution. We consider local autoregressive models, which predict each measured bit from up to $R$ preceding bits, so the context radius $R$ serves as the capacity parameter $s$. Their representation bias is exactly the summed conditional mutual information discarded beyond this radius. For a learner that estimates conditional probabilities using smoothed histograms, we derive an exact finite-sample estimation cost. When all contexts are well sampled, it reduces to the parameter-count law $V(R,N)\simeq d_R/(2N)$, giving $C(R)=2/d_R$, where $d_R$ counts the independent conditional-probability parameters.

Our target distributions are the computational-basis Born distributions of transverse-field Ising ground states. We independently evaluate the bias and estimation cost and use them to predict the observed optimal context radius across the tested couplings and sample sizes. In the gapped regimes, approximately exponential bias decay and exponentially growing parameter count give the row-I prediction of power-law decay of the optimal error in $N$. At criticality, algebraic decay of the infinite-chain bias instead gives the row-III prediction of asymptotic inverse-power decay in $\ln N$. Numerically measured optimal errors for bulk-site prediction are consistent with the corresponding finite-sample predictions.

\section{Bias--variance competition and scaling templates}
\label{sec:framework}

Let $p(x)=\operatorname{Tr}(\rho M_x)$ be the Born distribution of a state $\rho$ under a fixed measurement $\{M_x\}$, with outcomes $x\in\{0,1\}^n$. The training set $\XN=(x^{(1)},\dots,x^{(N)})\sim p^{\otimes N}$ consists of $N$ independent measurement records. A capacity parameter $s$ indexes model classes $\mathcal H_s$. We define the best population model and its representation bias by
\begin{equation}
q_s^\star\in\arg\min_{q\in\mathcal H_s}\KL{p}{q},
\qquad
B(s):=\KL{p}{q_s^\star}.
\label{eq:Bdef}
\end{equation}
For a model $\hat q_s=\hat q_s(\XN)\in\mathcal H_s$ learned from the training set, the excess population error
\begin{equation}
V_{\XN}(s):=\KL{p}{\hat q_s}-B(s)\ge0.
\label{eq:Vdef}
\end{equation}
Thus the decomposition
\begin{equation}
\KL{p}{\hat q_s}=B(s)+V_{\XN}(s)
\label{eq:decomp}
\end{equation}
is exact by definition. Averaging over training sets gives the mean estimation excess $V(s,N):=\E_{\XN}V_{\XN}(s)$ and the expected
total error $\mathcal E(s,N)=B(s)+V(s,N)$. We refer to $V$ as the variance term, although it includes any bias introduced by the learner and is not a literal estimator variance.

The optimal capacity and error are
\begin{equation}
\begin{aligned}
s^\star(N)&\in\arg\min_s\bigl[B(s)+V(s,N)\bigr],\\
\mathcal E^\star(N)&=\min_s\bigl[B(s)+V(s,N)\bigr].
\end{aligned}
\label{eq:rstar-def}
\end{equation}
For nested model classes, the representation bias $B(s)$ is nonincreasing with capacity. We focus on regimes where increasing $s$ reduces the bias but raises the estimation cost at fixed $N$, so their competition can produce an interior optimum. 

All information quantities are in nats. For comparisons across system sizes we use $b=B/n$, $v=V/n$, and $f=b+v$. This normalization leaves the optimum unchanged.

\subsection{Asymptotic estimation cost and scaling templates}
\label{sec:templates}

The asymptotic expansion in Eq.~\eqref{eq:law} holds at fixed capacity. Section~\ref{sec:finiteN:reg} proves it for the histogram learner, with $C(R)=2/d_R$. When $B(s)$ and $C(s)$ decrease with capacity, minimizing the asymptotic approximation to the total error, $B(s)+[NC(s)]^{-1}$, gives the interior stationarity condition
\begin{equation}
-\frac{\mathrm d B(s)}{\mathrm ds}
=
-\frac{1}{N\,C(s)^2}\frac{\mathrm d C(s)}{\mathrm ds}.
\label{eq:balance}
\end{equation}
The optimum $s^\star(N)$ therefore balances the marginal reduction in bias against the marginal increase in estimation cost. Combining exponential and algebraic forms for $B$ and $C$ yields the four scaling templates in Table~\ref{tab:templates}, derived in
Appendix~\ref{app:templates}.

The optimal capacity grows with sample size, so the asymptotic approximation must remain accurate over the capacities competing for the optimum. At finite sample sizes, rarely observed contexts can shift the optimum away from the asymptotic prediction. The exact finite-sample expression derived in Section~\ref{sec:finiteN:exact} accounts for these contexts, allowing us to predict the optimum and assess the accuracy of the asymptotic approximation.

\subsection{Relation to prior work}

Previous work has developed neural representations and tomography methods for learning quantum measurement statistics from data~\cite{Carleo2017,CarrasquillaTorlai2021,Torlai2018,Carrasquilla2019,Rocchetto2018,Quek2021}, including autoregressive models with normalized likelihoods and direct sampling~\cite{WuWangZhang2019,Sharir2020,HibatAllah2020}. Other approaches, such as classical shadows, target the estimation of selected observables~\cite{Huang2020}. Our focus is the choice of model capacity: for a specified measurement distribution and learner, we determine how much capacity a finite number of samples justifies.

Studies of neural scaling laws relate loss to dataset size and model size~\cite{Kaplan2020,Bahri2024}. For natural language, data-limited exponents have been related to pairwise correlations and the convergence of finite-context conditional entropy to its limiting value~\cite{Cagnetta2026_ScalingLawsNaturalLanguage}. In our framework, this entropy excess has a direct interpretation: for a stationary process, it equals the per-site representation bias of a context-limited autoregressive model (Lemma~\ref{lem:cmi}). We combine its dependence on capacity with the estimation cost to predict the sample-size dependence of the optimal error. In particular, row I shows that exponentially
decaying bias and exponentially growing estimation cost together produce power-law decay of the optimal error with sample size. Such power-law decay is therefore not, by itself, a signature of critical data.

In quantum systems, prior work has used conditional mutual information and related locality measures to characterize representability in different
phases~\cite{Kumar2026_UnlearnablePhases,Lu2026_InfoScalingNQS,ZhangGopalakrishnan2025_CMIGibbs,Yi2026_UniversalCMIDecay}. Complementary numerical work on simulated Rydberg-array measurements has found approximate power-law error scaling near criticality for a fixed-size autoregressive transformer, with less stable scaling away from it~\cite{Berman2026_QuantumModelsScale}. In our analysis, the exact relation between autoregressive representation bias and summed conditional mutual information connects the structure of the measurement distribution to the error introduced by limiting the context radius. We combine this bias with the estimation cost to predict how both the optimal radius and the optimal error change with sample size.

Determining this estimation cost requires specifying how the model is learned from finite data. Prior studies of discrete distribution estimation have analyzed the finite-sample KL risk of smoothed histogram estimators~\cite{KrichevskyTrofimov1981,BraessSauer2004,Paninski2004,Kamath2015}. Building on this work, we use a histogram learner with Krichevsky--Trofimov (KT) smoothing~\cite{KrichevskyTrofimov1981} for each context in a local autoregressive model. For this learner, we derive an exact finite-sample expression for the mean estimation excess at fixed radius. The expression accounts for rarely observed and unseen contexts, allowing us to evaluate the bias--variance tradeoff beyond the well-sampled approximation.

\section{Local autoregressive models}
\label{sec:localAR}

\subsection{Representation bias as discarded predictive information}

Order the sites $i=1,\ldots,n$. A radius-$R$ autoregressive model predicts each bit from at most
$R$ preceding bits. Write the retained context and discarded past as
\begin{equation}
\ctx=X_{\max(1,i-R):i-1},
\qquad \dpast=X_{<i-R},
\label{eq:context}
\end{equation}
where $\dpast$ is empty for $i\le R+1$, and the context is empty at $i=1$. Writing $c_{i,R}$ for a realization of $\ctx$, the model class is
\begin{equation}
\mathcal H_R^{\mathrm{loc}}=
\left\{q:q(x)=\prod_{i=1}^{n}q_i(x_i\mid c_{i,R})\right\},
\label{eq:class}
\end{equation}
where each conditional is an arbitrary normalized probability table. The context radius $R$ therefore serves as the capacity parameter. This class isolates the representation error caused by context truncation, without imposing additional restrictions on the conditional probabilities.

\begin{lemma}[Representation identity]
\label{lem:cmi}
For any target distribution $p$, a KL-optimal model in $\mathcal H_R^{\mathrm{loc}}$ is $q_R^\star(x)=\prod_i p(x_i\mid c_{i,R})$, and
\begin{equation}
B(R)=\KL{p}{q_R^\star}
=\sum_{i=1}^{n}I\bigl(X_i:\dpast\mid\ctx\bigr).
\label{eq:cmi}
\end{equation}
\end{lemma}

\begin{proof}
The KL chain rule separates the optimization over sites and contexts. For each context $c$, the minimizing conditional is $p(X_i\mid c)$. Substitution gives $\sum_i[H(X_i\mid\ctx)-H(X_i\mid X_{<i})]$. Since $X_{<i}=(\dpast,\ctx)$, this is Eq.~\eqref{eq:cmi}.
\end{proof}

Each term measures the additional information that the omitted past provides about the predicted bit, given the context. Thus $B(R)$ is nonnegative and nonincreasing with $R$, ranging from the
multi-information $B(0)=\sum_i H(X_i)-H(X)$ to $B(n-1)=0$. Its decay quantifies how accurately a local autoregressive model can
represent the target distribution. The representation identity itself holds for any distribution, including the quantum measurement distributions studied below.

\subsection{TFIM benchmark}
\label{sec:benchmark}

We study computational-basis Born distributions of ground states of the one-dimensional transverse-field Ising model
(TFIM)~\cite{Pfeuty1970},
\begin{equation}
H(g)=-\sum_{i=1}^{n-1}\sigma_i^z\sigma_{i+1}^z-g\sum_{i=1}^{n}\sigma_i^x,
\label{eq:tfim}
\end{equation}
with the Ising coupling set to one. For a finite open chain, the target distribution is $p_g(x)=|\braket{x|\psi_g}|^2$, where $\ket{\psi_g}$
is the ground state of $H(g)$. Autoregressive predictions follow the site ordering along the chain. We consider the ordered ($g<1$), critical ($g=1$), and disordered ($g>1$) regimes.

We use two complementary settings. The main study considers prediction at a bulk site of the infinite chain, avoiding boundary effects that truncate the predictive information in the past. This is especially important at criticality, where that information decays slowly with context radius. Each training record is an independently sampled block, from which the learner predicts one target bit using its $R$ preceding spins. The expected error relative to prediction from the full past is $f_g(R,N)=b_g(R)+v_g(R,N)$. This setting lets us test the predicted scaling of the optimal radius and error with sample size.

To test capacity selection for a complete measurement distribution, we also study a 50-site open chain. Here the error is the total KL divergence of the learned distribution, including contributions from boundary sites (Appendix~\ref{app:learner}).
\section{Representation bias across the TFIM phase diagram}
\label{sec:tfim}

For the stationary infinite chain, let $H_R$ be the entropy of $R$ consecutive spins, $h_R=H_{R+1}-H_R$ the one-step conditional entropy, and $h_\infty=\lim_{R\to\infty}h_R$ the entropy rate~\cite{CrutchfieldFeldman2003}. By Lemma~\ref{lem:cmi}, the bulk representation bias is $b_g(R)=h_R-h_\infty$, where the entropies depend on $g$. It measures the predictive information lost by
restricting the context to $R$ spins.

\begin{figure}[t]
\includegraphics[width=\columnwidth]{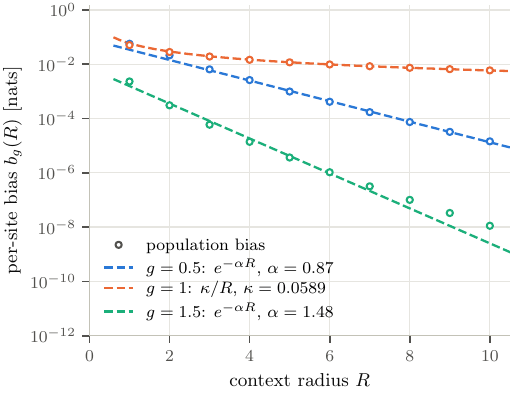}
\caption{Bulk representation bias $b_g(R)$ for $g=0.5$, $1$, and $1.5$. Circles show numerically computed values. Dashed curves show exponential guides in the gapped regimes and $\kappa/R$ at criticality, with $\kappa=0.0589$ nats from Ref.~\cite{LauGrassberger2013} and no fitted parameter. The exponential decay rates are averaged over the radii visited by the bulk optimizer. The uncertainty in the critical entropy rate is about $10^{-4}$ nats, smaller than the symbols. Numerical methods and details of the exponential guides are
given in Appendix~\ref{app:infinite}.}
\label{fig:bias}
\end{figure}

Figure~\ref{fig:bias} shows that the bias decreases with context radius in both gapped and critical regimes, but at different rates. At $g=0.5$ and $1.5$, the decay is approximately exponential over the resolved radii. At criticality ($g=1$), logarithmic corrections to Ising row
entropies~\cite{LauGrassberger2013} and to Shannon mutual information in quantum chains~\cite{AlcarazRajabpour2013,Stephan2014} suggest
an algebraic decay. If $H_R=h_\infty R+\kappa\ln R+O(1)$, with subleading terms whose successive differences are $o(R^{-1})$, then
\begin{equation}
b_{g=1}(R)=h_R-h_\infty\sim\frac{\kappa}{R}.
\label{eq:bulk-bias}
\end{equation}
The computed bias at $g=1$ is consistent with this prediction.

\begin{figure*}[t]
\includegraphics[width=\textwidth]{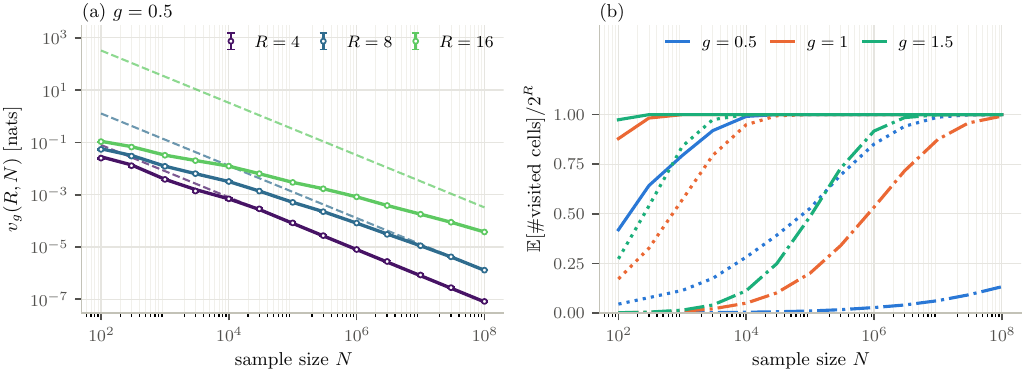}
\caption{Finite-sample estimation cost and context occupancy at a bulk site of the infinite chain. (a)~Mean estimation excess $v_g(R,N)$ at $g=0.5$ for $R=4$, $8$, and $16$. Points show averages over 20 training sets, with standard errors smaller than the symbols. The exact prediction
[Eq.~\eqref{eq:exact}, solid lines] agrees with the measurements, while the leading asymptotic estimation cost $2^R/(2N)$ (dashed lines) agrees
closely for $N\approx10^4$ at $R=4$, only at the largest sample sizes at $R=8$, and substantially overestimates the cost at $R=16$, where contexts remain poorly sampled. (b)~Expected fraction of contexts observed, $\sum_c[1-(1-\pi_c)^N]/2^R$, for $g=0.5$, $1$, and $1.5$ at $R=4$ (solid), $8$ (dotted), and $16$ (dash-dotted). Larger contexts require more samples to achieve comparable coverage. Context occupancy helps explain the departure from the parameter-count law in (a), although the exact cost also depends on the conditional probabilities within each context.}
\label{fig:mechanism}
\end{figure*}

Finite chains limit the available past and modify the bias, particularly at criticality (Appendix~\ref{app:infinite}). We therefore use the bulk bias $b_g(R)$ for the main learning study and evaluate the finite-chain bias $B_g(R)$ separately for the 50-site study in Appendix~\ref{app:learner}.

\section{Finite-sample estimation cost}
\label{sec:finiteN}

\subsection{The well-sampled regime: a parameter count}
\label{sec:finiteN:reg}

For each site--context pair $(i,c)$, define the context probability $\pi_{i,c}=\Pr(C_{i,R}=c)$ and the conditional parameter $\theta_{i,c}=\Pr(X_i=1\mid C_{i,R}=c)$. We refer to each pair as a cell. The histogram learner estimates these parameters separately using KT smoothing (Appendix~\ref{app:cell}), which keeps the KL error finite even when an outcome is absent from the training data.

For a fixed conditional parameter $0<\theta<1$, the mean KL error after $m$ visits is $1/(2m)+O(m^{-2})$ as $m\to\infty$. A cell with probability $\pi$ receives about $N\pi$ visits, so its probability-weighted contribution is $\pi/(2N\pi)=1/(2N)$ to leading order.
Thus each independent Bernoulli parameter contributes the same leading estimation cost.

\begin{corollary}[Leading asymptotic estimation cost]
\label{cor:dR}
For fixed $n$ and $0\le R\le n-1$, with $\pi_{i,c}>0$ and $0<\theta_{i,c}<1$ in every cell,
the histogram learner satisfies
\begin{equation}
\begin{aligned}
V(R,N)&=\frac{d_R}{2N}+O(N^{-2}),\\
d_R&=(n-R+1)2^R-1.
\end{aligned}
\label{eq:dRlaw}
\end{equation}
The remainder depends on the fixed context and conditional probabilities.
\end{corollary}

The parameter count $d_R$ therefore gives $C(R)=2/d_R$. At a bulk site of the infinite chain, there are $2^R$ context cells, giving $v(R,N)=2^R/(2N)+O(N^{-2})$ at fixed $R$. For a finite chain, counting the conditional parameters across all sites gives $d_R$ in Eq.~\eqref{eq:dRlaw}.

All finite-chain TFIM targets considered here have full support. At fixed $n$ and $R$, they share the same leading estimation cost, but their probabilities determine how many samples $N$ are needed to reach this regime. The parameter count grows exponentially with $R$ in the bulk and, for $R\ll n$, in the finite chain, placing these models in the exponential-$C$ rows of Table~\ref{tab:templates}. Appendix~\ref{app:exact} proves the corollary and derives the fixed-radius asymptotic expansion in integer powers of $1/N$.

\subsection{Exact mean estimation cost at finite sample size}
\label{sec:finiteN:exact}

At finite $N$, many cells can be rarely sampled. The expected visit count $N\pi_{i,c}$
distinguishes well-populated from sparse cells. Large visit counts are necessary for the
asymptotic approximation, but nearly deterministic conditionals can delay convergence further.
Averaging over both the visits and their outcomes gives the finite-sample estimation cost:

\begin{figure*}[t]
\includegraphics[width=\textwidth]{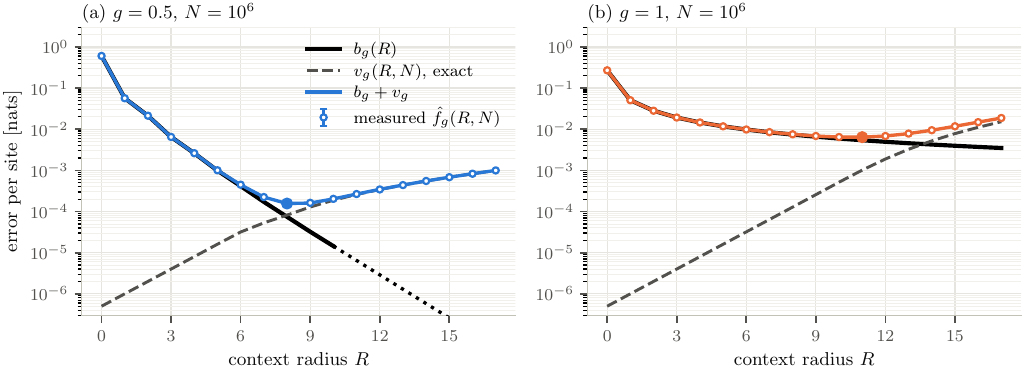}
\caption{Context-radius selection at a bulk site of the infinite chain for $N=10^6$, at (a)~$g=0.5$ and (b)~$g=1$.
Curves show the representation bias $b_g(R)$ (black), the exact mean estimation excess $v_g(R,N)$ (dashed), and their sum (colored). Points show the mean total error over 20 training sets, with standard errors smaller than the symbols. Filled markers indicate the predicted minima, which coincide with the measured minima. The dotted bias curve in (a) is a geometric extrapolation beyond $R=10$, where numerical precision limits the entropy calculation
(Appendix~\ref{app:infinite}). In (b), uncertainty in the entropy rate shifts the bias and total error equally at all radii, leaving the optimum unchanged.}
\label{fig:rstar}
\end{figure*}

\begin{proposition}[Exact mean estimation excess]
\label{prop:exact}
For $N$ i.i.d. full configurations and site-dependent tables,
\begin{equation}
\begin{aligned}
V(R,N)&=\sum_{i,c}\pi_{i,c}\,
\E_{M\sim\mathrm{Bin}(N,\pi_{i,c})}\,
\E_{K\sim\mathrm{Bin}(M,\theta_{i,c})}\\
&\hspace{9mm}\times
\KL{\Bern(\theta_{i,c})}{\Bern(\hat\theta_{M,K})},
\end{aligned}
\label{eq:exact}
\end{equation}
where $\hat\theta_{m,k}=(k+1/2)/(m+1)$ for KT. The identity also holds for any deterministic
cell estimator $\hat\theta_{m,k}\in(0,1)$.
\end{proposition}

The KL chain rule expresses the realized excess as
\begin{equation}
V_{\XN}(R)=\sum_{i,c}\pi_{i,c}
\KL{\Bern(\theta_{i,c})}{\Bern(\hat\theta_{i,c})}.
\label{eq:percell}
\end{equation}
Each site--context pair is encountered at most once per configuration, giving the binomial
counts in Eq.~\eqref{eq:exact}. Linearity of expectation completes the argument, and
correlations between sites and cells are allowed. Appendix~\ref{app:exact} gives the explicit sums and
numerical evaluation method.

Figure~\ref{fig:mechanism}(a) shows agreement between the exact prediction and the measured mean estimation excess for learners trained on independent bulk samples. Each learned model's excess is evaluated using Eq.~\eqref{eq:percell}, with further validation in Appendix~\ref{app:exact}.

Comparing $R=4$, $8$, and $16$ shows that increasing the context radius delays the well-sampled regime: lower context coverage
[Fig.~\ref{fig:mechanism}(b)] accompanies a slower approach to the parameter-count law. This effect is especially strong in the ordered regime, where context probabilities are highly uneven. The finite-chain study shows similar behavior (Appendix~\ref{app:learner}).

\section{Selecting the context radius}
\label{sec:rstar}

\subsection{Bias--variance competition at a bulk site}
\label{sec:selection}

We predict the optimal context radius by minimizing the sum of the independently evaluated bias and estimation cost:
\begin{equation}
R^\star_{\rm exact}(N)\in
\arg\min_R\left[b_g(R)+v_g(R,N)\right],
\label{eq:rstar-exact}
\end{equation}
using Eq.~\eqref{eq:cmi} for the bias and Eq.~\eqref{eq:exact} for the mean estimation excess.

Figure~\ref{fig:rstar} shows this competition at $N=10^6$.
At $g=0.5$, the rapidly decreasing bias and increasing estimation cost produce a sharp minimum at $R=8$. At criticality ($g=1$), the slower bias decay produces a broader minimum at $R=11$, where bias remains the larger contribution. In both panels, the measured mean total error agrees with the predicted sum.

Across the tested couplings and sample sizes, the exact
finite-sample cost predicts the measured optimal radius in all but one case, where the predicted and measured optima give mean errors that are indistinguishable within sampling uncertainty. The predicted optimal error also agrees with the measurements (Fig.~\ref{fig:scaling}). The asymptotic approximation $2^R/(2N)$ is less reliable, particularly at $g=0.5$: it overestimates the estimation cost and favors radii that are too small. The exact expression accounts for poorly sampled contexts and captures these shifts. Detailed comparisons and the corresponding finite-chain results are given in Appendix~\ref{app:learner}.

\subsection{Scaling at critical and gapped couplings}
\label{sec:critical}

\begin{figure*}[t]
\includegraphics[width=\textwidth]{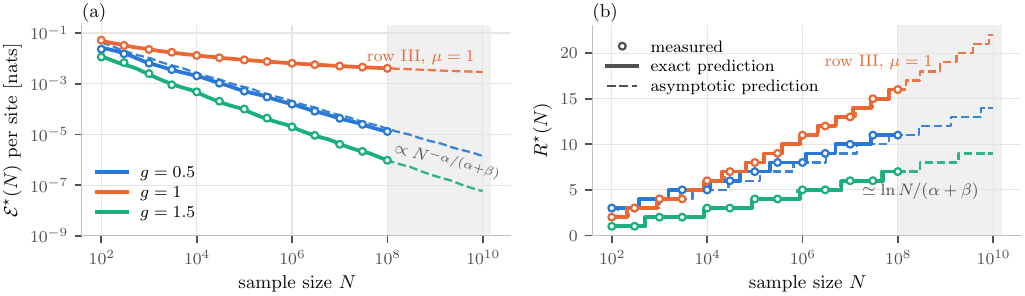}
\caption{Sample-size dependence of (a)~the optimal bulk-site error and (b)~the optimal context radius. Points show results from 20 training realizations. Solid curves are predictions from the computed bias and exact finite-sample estimation cost, without fitted parameters, over the range where the optimum lies within $R\le17$. Dashed curves minimize the asymptotic approximation to the total error over integer radii, using continued bias laws (Appendix~\ref{app:infinite}).
The shaded region beyond $N=10^8$ contains extrapolations only. At criticality, substantial subleading corrections persist over the sampled range, so the leading $1/\ln N$ behavior is not yet reached.}
\label{fig:scaling}
\end{figure*}

\begin{table*}[t]
\caption{Predicted and fitted scaling at a bulk site.
For the gapped couplings, $\alpha$ is the mean bias decay rate
over the radii visited by the optimizer, and the fitted entries
come from power-law error fits and logarithmic radius fits.
At criticality, $\mu=1$ follows from $b_g(R)\sim\kappa/R$.
The critical fitted entries are leading terms inferred from the
full row-III template, including subleading corrections.
They are not direct slopes of the data.
The leading asymptotic estimation cost gives $\beta=\ln2$ throughout.
Fit details are given in Appendix~\ref{app:templates}.}
\label{tab:exponents}
\setlength{\tabcolsep}{3pt}
\begin{ruledtabular}
\begin{tabular}{lcccccc}
$g$ (row) & bias & estimation cost
& \multicolumn{2}{c}{optimal error $f_g^\star(N)$}
& \multicolumn{2}{c}{optimal radius $R^\star(N)$}\\
& exponent & exponent & predicted & fitted & predicted & fitted\\
\colrule
$0.5$ (I)\rule{0pt}{2.6ex}
& $\alpha=0.87$ & $\beta=\ln2$
& $N^{-0.56}$ & $N^{-0.55}$
& $0.64\ln N$ & $0.63\ln N$\\
$1.5$ (I)
& $\alpha=1.48$ & $\beta=\ln2$
& $N^{-0.68}$ & $N^{-0.69}$
& $0.46\ln N$ & $0.43\ln N$\\
$1$ (III)
& $\mu=1$ & $\beta=\ln2$
& $(\ln N)^{-1}$ & $(\ln N)^{-1.01}$
& $1.44\ln N$ & $1.47\ln N$\\
\end{tabular}
\end{ruledtabular}
\end{table*}

Figure~\ref{fig:scaling} shows how the optimal error and radius depend on sample size. At the gapped couplings, the error decreases approximately as a power of $N$, while the optimal radius grows logarithmically, consistent with row I. At criticality, the error decreases much more slowly and the optimal radius is larger. The critical bias corresponds to the algebraic form $b(R)\propto R^{-\mu}$ with $\mu=1$. Using the template $b(R)=\kappa/R$ and the leading asymptotic estimation cost $2^R/(2N)$ gives
\begin{equation}
\begin{aligned}
R^\star(N)
&=\frac{\ln N}{\ln2}
-\frac{2\ln\ln N}{\ln2}+O(1),\\
f^\star_{g=1}(N)
&=\frac{\kappa\ln2}{\ln N}
+\frac{2\kappa\ln2\,\ln\ln N}{(\ln N)^2}
+O\!\left(\frac{1}{(\ln N)^2}\right).
\end{aligned}
\label{eq:critical-scaling}
\end{equation}
The first term in each expression gives the leading scaling, and the second gives the logarithmic correction.
These predictions require the asymptotic approximation to remain accurate near the optimum.

The relative correction to the optimal error, $2\ln\ln N/\ln N$, decays slowly and remains substantial over the sampled range. A joint fit of the optimal error and radius to the full row-III template, with the parameters free and subleading corrections retained, gives $\mu=1.01\pm0.03$, consistent with the predicted $\mu=1$
(Appendix~\ref{app:templates}). Table~\ref{tab:exponents} compares the predicted and fitted scaling of both quantities: the gapped entries describe effective scaling over the sampled range, while the critical entries are leading asymptotic terms inferred from the full template.

\section{Discussion}
\label{sec:discussion}

The bias--variance competition has an information-theoretic interpretation. The bias measures discarded conditional information, whereas the leading asymptotic estimation cost counts independently estimated conditional probabilities. For the site-dependent histogram learner,
\begin{equation}
\frac{1}{C(R)}
=\frac12\sum_i e^{H_0(C_{i,R})}
=\frac{d_R}{2},
\label{eq:information-dimension}
\end{equation}
where $H_0$ is the logarithm of the context support size, assuming $0<\theta_{i,c}<1$ for every supported context.

Different parts of the framework have different scopes. The decomposition in Eq.~\eqref{eq:decomp} applies to any capacity-indexed model family and learner. The CMI identity is exact for the full local autoregressive class. For a restricted parameterization, including a neural autoregressive model, it gives a lower bound on the representation bias. The exact estimation formula applies specifically to the KT histogram learner: it allows
correlations within each sample but assumes independent samples and separately estimated conditionals.

The predicted optimum is a population benchmark evaluated using the target probabilities available in these controlled examples. It therefore shows which
capacity is best in expectation, but does not provide a rule for selecting that capacity from the training data alone. Moreover, our results characterize the mean population error over training sets, rather than the error attained by an individual training set. High-probability guarantees require additional control of rare outcomes~\cite{VanDerHoeven2026}. PAC--Bayes methods provide a complementary route to such guarantees~\cite{McAllester1999,Catoni2007,Mbacke2023_VAE_PAC_Bayes,RodriguezGrasa2026_PACBayesQuantum}, although generic $N^{-1/2}$ bounds do not resolve the $N^{-1}$ leading behavior of the fixed-radius mean estimation excess.

The same capacity-selection problem arises in other architectures. Possible capacity parameters include the bond dimension of an MPS~\cite{Han2018_MPSGenerative,Glasser2019}, the receptive field of a local diffusion model~\cite{Hu2025_LocalDiffusionPhases}, and the retained dimension of a hierarchical representation~\cite{KochJanuszRingel2018}. For each architecture, the representation bias and estimation cost
must be established separately. Comparing architectures therefore requires optimizing their capacities at the same sample size and evaluating the resulting population errors. Accordingly, the capacity justified by finite data is a property of the measured distribution, model family, learner, and sample size, rather than of the quantum state alone.
\section{Conclusion}
\label{sec:conclusion}

Bias--variance competition connects the structure of quantum measurement data to the model capacity supported by finite data. For local autoregressive models, the representation bias is exactly the summed conditional mutual information lost by truncating the context. For the KT histogram learner, the mean estimation excess has an exact finite-sample expression that recovers the parameter-count law in the well-sampled regime and accounts for rarely observed contexts outside it. Applied to TFIM Born distributions, these independently evaluated quantities predict the optimal context radius and optimal error without fitting either optimum.

At a bulk site of the infinite chain, the bias decays approximately exponentially in the gapped regimes and algebraically at criticality. Combined with the exponentially growing asymptotic estimation cost, these behaviors predict power-law decay of the optimal error in $N$ and inverse-power decay in $\ln N$, respectively. The critical finite-sample results are consistent with the latter prediction when its substantial subleading corrections are retained. Therefore, power-law learning curves of the kind reported for neural learners~\cite{Cagnetta2026_ScalingLawsNaturalLanguage} therefore do not by themselves indicate critical data, and a logarithmic learning curve in~\cite{Hou2025_MLManyMeasurements} has a mechanism in algebraic bias decay against an exponentially growing estimation cost. 

Applying the framework to other architectures requires determining their representation-bias and estimation-cost laws. A practical next step is to replace the population benchmark with a capacity-selection rule that uses only the observed samples.

\begin{acknowledgments}
We thank Yu-Hsueh Chen, Ehud Altman, and John McGreevy for discussions of the critical conditional mutual information.
This work was partially supported by the U.S. Department of Energy under Award No.~DE-SC0026807.

The authors used ChatGPT (OpenAI, GPT-5.6 Sol) and Claude
(Anthropic, Claude Fable 5.1) to assist with code development and manuscript editing. All AI-assisted code was reviewed and tested, and all AI-assisted text was reviewed and revised by the authors. The authors take full responsibility for the code and scientific content of this work.
\end{acknowledgments}

\emph{Data and code availability.}---The data and code needed to reproduce the figures, tables, and reported numerical results are available at
\url{https://github.com/S00Zhang/quantum-measurement-sample-complexity}.

\appendix
\section{Finite-sample estimation theory}
\label{app:estimation}

\subsection{Per-cell KT error}
\label{app:cell}

Let $n_{i,c}$ count visits to cell $(i,c)$ and let $n_{i,c,1}$ count outcomes $X_i=1$ in that cell. The Krichevsky--Trofimov estimate is
\begin{equation}
\hat\theta_{i,c} = \frac{n_{i,c,1}+\tfrac12}{n_{i,c}+1}.
\label{eq:kt}
\end{equation}
It is the posterior-predictive mean for a $\mathrm{Beta}(1/2,1/2)$ prior and assigns probability $1/2$ to an unseen context~\cite{KrichevskyTrofimov1981,Jeffreys1946,ClarkeBarron1994}.

\begin{lemma}[Per-cell KT error]
\label{lem:cell}
For fixed $\theta\in(0,1)$ and $K\sim\mathrm{Bin}(m,\theta)$, define
\begin{equation}
r_m(\theta) = \E\, \KL{\Bern(\theta)}{\Bern(\hat\theta_{\rm KT})}.
\end{equation}
Then
\begin{equation}
r_m(\theta) = \frac{1}{2m} + \left[ \frac{1}{24\theta(1-\theta)} - \frac{5}{12} \right] \frac{1}{m^2} + O(m^{-3}).
\label{eq:cell-law}
\end{equation}
\end{lemma}

\begin{proof}
Let $S=\theta(1-\theta)$, $u=\tfrac12-\theta$, and $\Delta=\hat\theta_{\rm KT}-\theta$, where $\hat\theta_{\rm KT}=(K+\tfrac12)/(m+1)$. Expanding the Bernoulli KL divergence about $\Delta=0$ gives
\begin{equation}
\begin{aligned}
\KL{\Bern(\theta)}{\Bern(\theta+\Delta)} &= \frac{\Delta^2}{2S} + \frac{2\theta-1}{3S^2}\Delta^3\\
&\quad + \frac{1-3S}{4S^3}\Delta^4 + O(\Delta^5).
\end{aligned}
\end{equation}
The corresponding binomial moments are
\begin{equation}
\begin{aligned}
\E[\Delta^2] &= \frac{S}{m} + \frac{u^2-2S}{m^2} + O(m^{-3}),\\
\E[\Delta^3] &= \frac{5uS}{m^2} + O(m^{-3}),\\
\E[\Delta^4] &= \frac{3S^2}{m^2} + O(m^{-3}).
\end{aligned}
\end{equation}
Substitution, with $u^2=\tfrac14-S$, gives Eq.~\eqref{eq:cell-law}. The Taylor remainder is controlled on a fixed neighborhood of $\theta$, while the contribution outside that neighborhood is exponentially small by binomial concentration. Its expectation is therefore $O(m^{-3})$.
\end{proof}

The expansion holds at fixed $\theta$. Its second-order coefficient diverges as $\theta\to0,1$, so nearly deterministic conditionals require larger occupancies to reach the well-sampled regime.

\subsection{Exact identity and fixed-radius expansion}
\label{app:exact}
\label{app:expansion}

Each independent training configuration visits a site--context cell at most once. Hence $M_{i,c}\sim\mathrm{Bin}(N,\pi_{i,c})$ and $K_{i,c}\mid M_{i,c}=m\sim\mathrm{Bin}(m,\theta_{i,c})$. Averaging the cellwise KL decomposition in Eq.~\eqref{eq:percell} gives Proposition~\ref{prop:exact}, without requiring independence across cells. Explicitly,
\begin{equation}
\begin{aligned}
V(R,N) = \sum_{i,c}\pi_{i,c}\sum_{m=0}^{N} &\binom Nm \pi_{i,c}^{m}(1-\pi_{i,c})^{N-m}\\
{}\times\sum_{k=0}^{m} &\binom mk \theta_{i,c}^{k}(1-\theta_{i,c})^{m-k}\\
{}\times& \KL{\Bern(\theta_{i,c})}{\Bern(\hat\theta_{m,k})},
\end{aligned}
\label{eq:exact-explicit}
\end{equation}
where $\hat\theta_{m,k}=(k+\tfrac12)/(m+1)$. In particular, an unvisited cell is assigned $\hat\theta_{0,0}=1/2$.

The parameter-count law of Corollary~\ref{cor:dR} is the leading term of an asymptotic expansion of each weighted cell contribution. Define
\begin{equation}
v_N(\pi,\theta) = \pi\, \E_{M\sim\mathrm{Bin}(N,\pi)} r_M(\theta),
\label{eq:vN}
\end{equation}
so that
\begin{equation}
V(R,N) = \sum_{i,c} v_N(\pi_{i,c},\theta_{i,c}).
\end{equation}

\begin{proposition}[Large-$N$ expansion of the weighted cell error]
\label{prop:expansion}
Fix $0<\pi\le1$ and $0<\theta<1$. For every integer $J\ge1$,
\begin{equation}
v_N(\pi,\theta) = \sum_{j=1}^{J} \frac{c_j(\pi,\theta)}{N^j} + O(N^{-J-1}),
\label{eq:expansion}
\end{equation}
where the coefficients and implied constant depend on $\pi$, $\theta$, and $J$. The first two coefficients are
\begin{equation}
\begin{aligned}
c_1 &= \frac12,\\
c_2 &= \frac1\pi \left[ \frac{1-\pi}{2} + \frac{1}{24\theta(1-\theta)} - \frac5{12} \right].
\end{aligned}
\label{eq:c1c2}
\end{equation}
\end{proposition}

Summing Eq.~\eqref{eq:expansion} over the fixed set of $d_R$ cells proves Corollary~\ref{cor:dR}. It also shows that, at fixed radius, the mean excess contains no fractional powers of $N$, including no $N^{-1/2}$ term. The unweighted error of a cell is $1/(2N\pi)+O(N^{-2})$: a rare cell is learned more slowly, but its probability weight $\pi$ cancels this factor in the leading contribution.

\begin{proof}
We first expand the error at fixed visit count. Write $\Delta=\hat\theta_{m,K}-\theta=(Z+u)/(m+1)$, where $Z=K-m\theta$, $K\sim\mathrm{Bin}(m,\theta)$, and $u=\tfrac12-\theta$. Taylor expansion of the Bernoulli KL divergence about $\Delta=0$, together with the fact that the central moments $\E Z^\ell$ are polynomials in $m$, gives
\begin{equation}
r_m(\theta) = \sum_{j=1}^{J} \frac{a_j(\theta)}{m^j} + O(m^{-J-1}).
\label{eq:rm-expansion}
\end{equation}
The expansion contains only integer powers of $1/m$. Its remainder is controlled on a fixed neighborhood of $\theta$, while the contribution outside that neighborhood is exponentially small by binomial concentration. Lemma~\ref{lem:cell} gives
\begin{equation}
a_1=\frac12, \qquad a_2 = \frac{1}{24\theta(1-\theta)} - \frac5{12}.
\end{equation}

Now let $M\sim\mathrm{Bin}(N,\pi)$ and define $A_N=\{M\ge N\pi/2\}$. The complement has exponentially small probability, while the KT loss grows at most logarithmically with $N$, so its contribution to $v_N$ is smaller than any power of $1/N$. On $A_N$, write $M=N\pi+W$ and expand each inverse power $M^{-j}$ about $N\pi$. Since the central moments of $W$ are polynomials in $N$, their expectations produce only integer powers of $1/N$. This gives Eq.~\eqref{eq:expansion}.

The leading contribution is $\pi a_1/(N\pi)=1/(2N)$. At second order,
\begin{equation}
\E\bigl[M^{-1};A_N\bigr] = \frac{1}{N\pi} + \frac{1-\pi}{N^2\pi^2} + O(N^{-3}).
\end{equation}
Combining this expression with $\pi a_2/(N\pi)^2$ gives the coefficient $c_2$ in Eq.~\eqref{eq:c1c2}.
\end{proof}

Equation~\eqref{eq:expansion} is an asymptotic expansion and does not assert convergence of the corresponding power series. It holds at fixed $\pi$ and $\theta$, and hence at fixed radius $R$. It does not constrain the powers appearing after optimization, where $R^\star(N)$ grows with $N$. The expansion is also nonuniform as $\pi\to0$ or $\theta\to0,1$, explaining why rare contexts and nearly deterministic conditionals require larger samples to reach the well-sampled regime. The exact identity in Eq.~\eqref{eq:exact-explicit} does not require these asymptotic conditions.

\emph{Context occupancy.} The expected number of contexts observed at site $i$ in $L$ independent configurations is
\begin{equation}
\sum_c \left[ 1-(1-\pi_{i,c})^L \right],
\label{eq:expected-occupancy}
\end{equation}
the quantity used in Fig.~\ref{fig:mechanism}(b) with $L=N$. It depends on the full distribution of context probabilities and approaches the number of supported contexts, $e^{H_0(i)}$, as $L\to\infty$. Occupancy alone does not determine the estimation cost, which also depends on the conditional parameters $\theta_{i,c}$ through Eq.~\eqref{eq:exact}.

\subsection{Numerical evaluation}

We evaluate Eq.~\eqref{eq:exact} using binomial sums for the per-cell KT error and then average over the visit count. At large occupancies, we use the expansion in Eq.~\eqref{eq:cell-law}. Near deterministic conditionals, we retain the numerical binomial evaluation. Grid refinement verifies numerical convergence. Comparisons with independent training realizations are shown for the bulk-site and finite-chain studies in Figs.~\ref{fig:mechanism}, \ref{fig:scaling}, \ref{fig:finite-chain-validation}, and \ref{fig:finite-chain-scaling}.

\section{TFIM population calculations}
\label{app:population}

\subsection{Free-fermion probabilities and finite-chain bias}
\label{app:popbias}

The ground state of the open TFIM chain is Gaussian in the Jordan--Wigner Majorana fermions. Its covariance matrix determines the spin moments through Pfaffians, and a Walsh--Hadamard transform gives the required window probabilities. For $g>0$, the finite-chain ground state has strictly positive computational-basis amplitudes, so all cells satisfy the support conditions used in Corollary~\ref{cor:dR}.

For the $50$-site calculation, we evaluate windows containing the target spin and up to $W=12$ preceding spins. These distributions give the cells $(\pi_{i,c},\theta_{i,c})$ and the conditional entropies $H(X_i\mid C_{i,R})$ for $R\le W$. We write the full-past conditional entropy as
\begin{equation}
\begin{aligned}
H(X_i\mid X_{<i}) &= H(X_i\mid C_{i,W})-T_i,\\
T_i &= \E_{\rm ctx} \KL{p(x_i\mid x_{<i})}{p(x_i\mid C_{i,W})}.
\end{aligned}
\label{eq:tail}
\end{equation}
The tail $T_i$ vanishes for $i\le W+1$ and is otherwise evaluated by averaging exact conditional divergences over contexts generated by sequential Gaussian conditioning. Its numerical uncertainty enters every $B_g(R)$ with $R\le W$ as a common additive offset and therefore does not affect radius selection. Comparisons with exact diagonalization on small chains validate the cell probabilities and tail-corrected bias.

\subsection{Conditional information of the infinite chain}
\label{app:infinite}

For $\ell$ consecutive bulk sites, the even-spin $\sigma^z$ expectation values are Pfaffians of the corresponding Majorana two-point matrices. A Walsh--Hadamard transform gives the $2^\ell$ block probabilities and their entropy $H_\ell$. We compute up to $\ell_{\max}=22$ at $g=0.5$ and $20$ at $g=1$ and $1.5$, giving $h_R=H_{R+1}-H_R$ for $R<\ell_{\max}$.

At the gapped couplings, we approximate $h_\infty$ by $h_{\ell_{\max}-1}$. The computed biases remain above the numerical precision floor over the radii used in the finite-$N$ predictions. The exponential guides in Fig.~\ref{fig:bias} use the mean decay rates $\alpha=0.87$ over $R=3$--$10$ at $g=0.5$ and $\alpha=1.48$ over $R=1$--$7$ at $g=1.5$, corresponding to the radii visited by the optimizer. These effective rates vary with radius and do not establish the asymptotic tail. The extrapolations in Fig.~\ref{fig:scaling} continue the computed gapped biases geometrically beyond the resolved range. The measured optima and exact finite-$N$ predictions use the computed values wherever they are available.

\begin{figure}[t]
\includegraphics[width=\columnwidth]{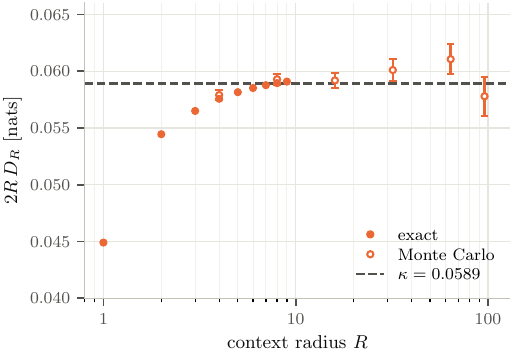}
\caption{Critical-tail test on the infinite chain at $g=1$. The compensated doubling difference $2R D_R$ of Eq.~\eqref{eq:DR}, in which $h_\infty$ cancels, is evaluated from exact block entropies (filled circles) and Monte Carlo averages of exact conditional divergences (open circles). The dashed line is the reference $\kappa=0.0589$ nats. If $b(R)\sim\kappa/R$, then $2R D_R\to\kappa$.}
\label{fig:critical-tail}
\end{figure}

At criticality, the two-point conditional mutual information
\begin{equation}
c_R = I(X_0:X_{-R-1}\mid X_{-R:-1}) = h_R-h_{R+1}.
\label{eq:two-point-cmi}
\end{equation}
is a second difference of the block entropy. The laws $c_R\sim\kappa/R^2$ and $b^{\rm bulk}(R)\sim\kappa/R$ are consistent because $b^{\rm bulk}(R)=\sum_{r\ge R}c_r$. To test the tail without first estimating $h_\infty$, we use the information gained by doubling the context,
\begin{equation}
D_R = h_R-h_{2R} = b^{\rm bulk}(R)-b^{\rm bulk}(2R).
\label{eq:DR}
\end{equation}
Under the algebraic law, $D_R\sim\kappa/(2R)$, and the bias can be reconstructed as $b(R)=\sum_{k\ge0}D_{2^kR}$.

Differences with $2R\le19$ follow from exact block entropies. At larger radii, sequential Gaussian conditioning gives the prediction from a finite spin context, allowing us to evaluate
\begin{equation}
D_R = \E_{\rm ctx} \KL{p(x\mid C_{2R})}{p(x\mid C_R)}.
\label{eq:DR-estimator}
\end{equation}
This Monte Carlo average is separate from the training experiments. The conditioning calculation agrees with exact diagonalization and Pfaffian block probabilities where both are available, and the compensated differences remain within $4\%$ of $\kappa=0.0589$ nats over the tested radii.

We approximate the residual tail by $b(128)=1.5D_{96}$, as implied by the $1/R$ law. This continuation becomes material at the largest radii, contributing up to $24\%$ of the reconstructed bias at $R=32$. It gives an uncertainty of about $10^{-4}$ nats in the inferred entropy rate. Combining this estimate with the exact $h_R$ supplies the critical biases used in the main text. For $R\le12$, they agree with $\kappa/R$ within $15\%$.

\subsection{Finite-chain bias and boundary effects}
\label{app:finitefits}

Open boundaries truncate the discarded past and suppress the critical contribution as the context approaches the available history. Site-resolved calculations on the $20$-site chain show approximately uniform conditional mutual information away from the ends, with boundary corrections to the site average. This truncation limits the critical scaling range of the finite chain and motivates the bulk-site study in the main text.

At the gapped couplings, the $50$-site bias is well described over the radii visited by the optimum by $B_g(R)/n\simeq A_g e^{-R/\xi_g}$. Together with the leading asymptotic estimation cost $d_R/(2N)\simeq(n-R+1)2^R/(2N)$, this gives the effective row-I prediction
\begin{equation}
R^\star(N) \simeq \frac{\ln N}{1/\xi_g+\ln2}+O(1),
\label{eq:loglaw}
\end{equation}
and $\mathcal E_g^\star(N)\asymp N^{-1/(1+\xi_g\ln2)}$. Fits over the radii visited by the optimum give $\xi_g=1.081$ at $g=0.5$ and $\xi_g=0.614$ at $g=1.5$. The resulting radius slopes and optimal-error exponents agree with the finite-chain results in Fig.~\ref{fig:finite-chain-scaling}. These are effective finite-range laws because the bias slope varies with radius and finite-size effects eventually become important.

\section{Learning protocols and finite-chain validation}
\label{app:learner}

\subsection{Bulk-site protocol}

The bulk-site study of Secs.~\ref{sec:finiteN} and~\ref{sec:rstar} uses $R=0,\ldots,17$ and 13 sample sizes from $10^2$ to $10^8$, with 20 independent training realizations. Each training record is an independent block of 18 consecutive bulk spins, drawn by inverse transform from its exact distribution. The final spin is the target, and its $R$ immediate predecessors form the context, so each record supplies one target--context observation. There is no pooling over sites. The target cells are marginals of the block distribution and are used both to score each fitted model and to evaluate Eq.~\eqref{eq:exact}. Every model uses the KT rule without a probability floor.

The bias uses the gapped entropy-rate approximation or the critical doubling-sum estimate of Appendix~\ref{app:infinite}. The latter introduces a common uncertainty of about $10^{-4}$ nats in the total error across radii, without affecting the optimum. The exact finite-$N$ prediction agrees with the measured mean excess throughout the tested grid and predicts the measured optimum except in one statistically unresolved case. The parameter-count law misses primarily in the ordered regime, where uneven context probabilities delay the onset of the well-sampled regime.

\subsection{Finite-chain protocol and validation}

The finite-chain study learns the complete Born distribution of the 50-site chain for $g=0.5$, $1$, and $1.5$, with $R=0,\ldots,12$ and nine sample sizes from $10^2$ to $10^6$. We use 20 independent training realizations. The error is the total KL divergence of a configuration, $B_g(R)+V_g(R,N)$, with contributions summed over sites and the bias evaluated as described in Appendix~\ref{app:popbias}. Within each realization, we generate the largest dataset by sequential Gaussian conditioning and use its prefixes for smaller $N$. The same samples are reused across radii. The population excess of each fitted model is evaluated from the exact cells through Eq.~\eqref{eq:percell}. The total error inherits the numerical uncertainty in $B_g$, which is common to all radii.

\begin{figure*}[t]
\includegraphics[width=\textwidth]{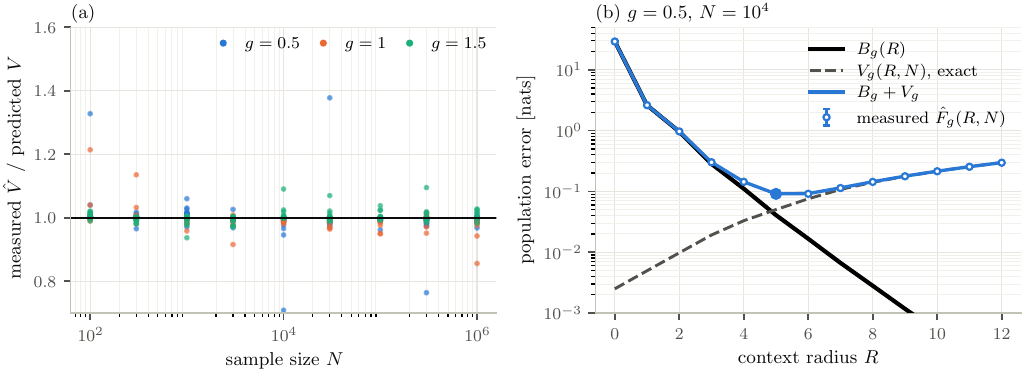}
\caption{Finite-chain validation and context-radius selection for the complete Born distribution of the $50$-site chain, using 20 training datasets. (a)~Ratio of the measured mean estimation excess to the exact finite-$N$ prediction across the tested grid. (b)~Representation bias $B_g(R)$, exact mean estimation excess $V_g(R,N)$, their sum, and the measured mean total error at $g=0.5$ and $N=10^4$. The filled marker indicates the predicted minimum, which coincides with the measured minimum.}
\label{fig:finite-chain-validation}
\label{fig:validation}
\end{figure*}

Figure~\ref{fig:finite-chain-validation}(a) confirms the exact mean cost across the tested finite-chain grid. Panel~(b) shows how this cost combines with the bias in one representative case: their sum is minimized at $R=5$ and follows the measured mean total error across radii.

\begin{figure*}[t]
\includegraphics[width=\textwidth]{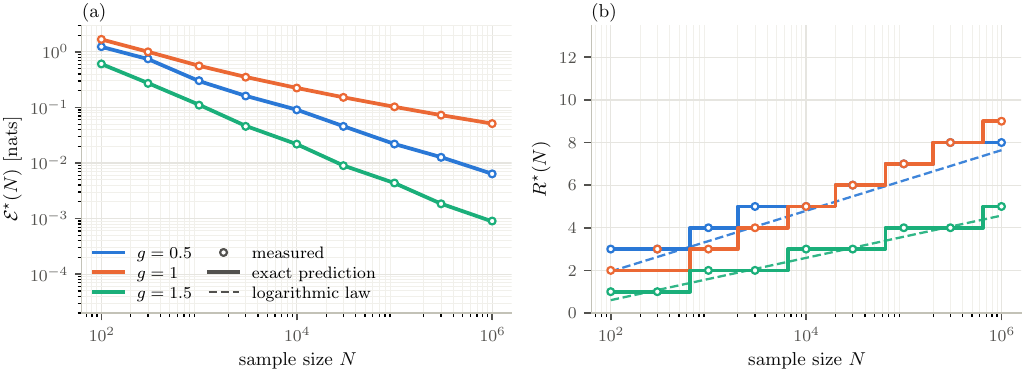}
\caption{Sample-size dependence for the complete Born distribution of the $50$-site chain. (a)~Measured optimal total error (points) and the exact finite-$N$ prediction (lines). (b)~Measured optimal context radii (points) and the exact finite-$N$ prediction (steps). Dashed curves in (b) show the gapped logarithmic law of Eq.~\eqref{eq:loglaw}, using bias parameters fitted independently of the optima. The measured quantities are obtained from mean errors over 20 training datasets.}
\label{fig:finite-chain-scaling}
\label{fig:rstar-n50}
\end{figure*}

Figure~\ref{fig:finite-chain-scaling} shows the optimal error and radius across sample sizes. The exact finite-$N$ prediction agrees with the measured optimum except at $g=1$ and $N=300$, where the two competing radii give errors that differ by less than one standard error. The parameter-count law is less accurate in the ordered regime, where many contexts remain unobserved. The gapped optima follow Eq.~\eqref{eq:loglaw}. At criticality, the finite available past limits the scaling range, motivating the bulk-site study of Sec.~\ref{sec:critical}.

\section{Derivation and fitting of the scaling templates}
\label{app:templates}

\subsection{Marginal balance and asymptotic precision}

Suppose the asymptotic approximation to the total error, $\mathcal E(s,N)\simeq B(s)+G(s)/N$ with $G=1/C$, remains accurate near the selected capacity. Treating $s$ as continuous gives $-B'(s)=G'(s)/N$. For the four templates below, positive multiplicative amplitudes shift $s^\star$ by $O(1)$ in rows I--III and change constant factors in row IV. We first set these amplitudes to one.

\emph{Row I:} $B=e^{-\alpha s}$ and $G=e^{\beta s}$. The balance condition gives $e^{(\alpha+\beta)s}=\alpha N/\beta$, and hence
\[
s^\star = \frac{\ln N}{\alpha+\beta}+O(1).
\]
Both costs scale as $N^{-\alpha/(\alpha+\beta)}$.

\emph{Row II:} $B=e^{-\alpha s}$ and $G=s^\nu$. The balance condition
\[
\alpha e^{-\alpha s} = \frac{\nu s^{\nu-1}}{N}
\]
implies $\alpha s=\ln N-(\nu-1)\ln s+O(1)$, giving
\[
s^\star = \frac{\ln N-(\nu-1)\ln\ln N}{\alpha}+O(1).
\]
At the optimum, the estimation cost $s^\nu/N$ exceeds the bias by a factor proportional to $s$, so $\mathcal E^\star\asymp(\ln N)^\nu/N$.

\emph{Row III:} $B=s^{-\mu}$ and $G=e^{\beta s}$. The balance condition
\[
\mu s^{-\mu-1} = \frac{\beta e^{\beta s}}{N}
\]
implies $\beta s=\ln N-(\mu+1)\ln s+O(1)$, giving
\[
s^\star = \frac{\ln N-(\mu+1)\ln\ln N}{\beta}+O(1).
\]
The estimation cost is $\mu/(\beta s^{\mu+1})$, smaller than the bias by a factor proportional to $1/s$, so $\mathcal E^\star\asymp(\ln N)^{-\mu}$.

\emph{Row IV:} $B=s^{-\mu}$ and $G=s^\nu$. The balance condition
\[
\mu s^{-\mu-1} = \frac{\nu s^{\nu-1}}{N}
\]
gives $s^\star\asymp N^{1/(\mu+\nu)}$, and both costs scale as $N^{-\mu/(\mu+\nu)}$.

The different logarithmic corrections in rows II and III follow from differentiation: $G'=\nu s^{\nu-1}$ lowers the power, whereas $-B'=\mu s^{-\mu-1}$ raises the inverse power. For the unit-amplitude templates, the balance condition also gives
\[
\mathcal E^\star = \frac{s^{\star\nu}}{N} \left[ 1+\frac{\nu}{\alpha s^\star} \right]
\]
in row II and
\[
\mathcal E^\star = s^{\star-\mu} \left[ 1+\frac{\mu}{\beta s^\star} \right]
\]
in row III. Substituting the refined optima yields
\begin{equation}
\begin{aligned}
\mathcal E^\star = {}& \frac{(\ln N)^\nu}{\alpha^\nu N}\\
&\times \left[ 1-\frac{\nu(\nu-1)\ln\ln N}{\ln N} + O\!\left(\frac{1}{\ln N}\right) \right]
\end{aligned}
\label{eq:row2-correction}
\end{equation}
and
\begin{equation}
\begin{aligned}
\mathcal E^\star = {}& \frac{\beta^\mu}{(\ln N)^\mu}\\
&\times \left[ 1+\frac{\mu(\mu+1)\ln\ln N}{\ln N} + O\!\left(\frac{1}{\ln N}\right) \right],
\end{aligned}
\label{eq:row3-correction}
\end{equation}
respectively.

Table~\ref{tab:templates} therefore states the capacities in rows I--III to additive $O(1)$ and the optimal errors up to bounded positive factors. At the latter precision, the relative $\ln\ln N/\ln N$ corrections need not be retained. The displayed prefactors in Eqs.~\eqref{eq:row2-correction} and~\eqref{eq:row3-correction} assume unit amplitudes.

This distinction concerns the value of the optimized error. It does not justify using only the leading capacity as a plug-in choice. In row III, choosing $s=\ln N/\beta$ gives $G(s)/N=1$, whereas the refined optimum makes this cost vanish. The $\ln\ln N$ term is therefore essential for locating a capacity that realizes the stated error law, even though its contribution to the optimal error is relatively small as $N\to\infty$.

\subsection{Fits to the bulk results}

At the gapped couplings, Table~\ref{tab:exponents} compares the row-I predictions with least-squares fits of $\ln\mathcal E^\star$ and $R^\star$ against $\ln N$ over the sampled range. These are effective exponents because the bias slope varies with radius and the asymptotic approximation is not uniformly accurate over this range.

At criticality, fits using only the leading forms over $10^2\le N\le10^8$ give an effective logarithmic exponent of $-1.8$ for the error and a radius slope of $1.0$, rather than the asymptotic predictions $-1$ and $1/\ln 2$. This discrepancy reflects the slowly decaying corrections derived above. We therefore fit the complete row-III surrogate
\begin{equation}
\mathcal E_{\rm T}(N) = \min_{R\ge1} \left[ \kappa R^{-\mu} + \frac{e^{\beta R}}{2N} \right]
\label{eq:critical-template}
\end{equation}
together with its minimizing integer radius $R_{\rm T}(N)$, as in the dashed curves of Fig.~\ref{fig:scaling}. A joint fit to the 13 measured optimal errors and radii, with all three parameters free, gives $\mu=1.01\pm0.03$, $\kappa=0.061\pm0.004$, and $\beta=0.680\pm0.006$, consistent with the reference values $1$, $0.0589$, and $\ln 2$. The fitted values are insensitive to varying the relative weight assigned to the radius residual over an order of magnitude.

The uncertainties are bootstrap standard errors obtained by resampling the 20 training realizations. They do not include systematic uncertainty from the bias continuation, asymptotic approximation, or template choice. The fit therefore tests consistency with the proposed critical law rather than independently establishing its asymptotic exponent. Outside the well-sampled regime, the exact estimation cost in Eq.~\eqref{eq:exact} must replace the template's estimation term.

\bibliography{bib/references}

\end{document}